# Origin of High-Temperature Antiferromagnetic Order in a van der Waals Material

Rabindra Basnet,[1] Hari Paudyal,[2] Gicela Saucedo Salas,[3]
Nicholas P. Butch,[3,4] Durga Paudyal,[2] and Ramesh C. Budhani[1,*]

[1]*Department of Physics, Morgan State University, Baltimore, MD, 21251, USA*
[2]*Department of Physics and Astronomy, University of Iowa, Iowa City, Iowa 52242, USA*
[3]*Maryland Quantum Materials Center, Department of Physics, University of Maryland, College Park, Maryland 20742, USA*
[4]*NIST Center for Neutron Research, 100 Bureau Drive, Gaithersburg, MD 20899-6102, USA*

## Abstract

While Van der Waals (vdW) itinerant antiferromagnets with high Néel temperatures ($T_N$) are highly desirable for spintronics, they remain relatively scarce. Here, we unravel the physical origin of the unusually high $T_N$ (≈ 250 K) in the newly identified vdW compound $(Fe_{0.65}Co_{0.35})_4GeTe_2$. The substitution of Co in $Fe_4GeTe_2$ induces layer-selective Fe/Co ordering and stabilizes a robust antiferromagnetic (AFM) state primarily driven by Co moments. The AFM order is further strengthened by enhanced electronic correlations of quasi–localized Co 3*d*-states at the Fermi level, giving rise to an itinerant–localized duality of the 3*d* electrons. This interplay generates strong magnetic correlations well above $T_N$ and stabilizes low–temperature spin canting with a possible nontrivial Berry curvature. Our results establish $(Fe_{0.65}Co_{0.35})_4GeTe_2$ as a rare material bridging fundamental magnetic interactions with potential applications in AFM spintronics.

[*]ramesh.budhani@morgan.edu

Van der Waals (vdW) antiferromagnets capitalize on the interplay between antiferromagnetic (AFM) order and two-dimensionality, offering negligible stray fields, ultrafast spin dynamics, and high crystalline quality for spintronics [1,2]. Metallic vdW antiferromagnets with high Néel temperatures ($T_N$) are particularly appealing [3,4] but remain rare due to intrinsic limitations from competing Coulomb interactions and electronic kinetic energy [5]. While magnetic coupling in localized systems is governed by the exchange mechanism [6], itinerant magnetism originates from the collective behavior of delocalized electrons. In the itinerant limit, the magnetic order is governed by the Stoner criterion, whereby spontaneous spin polarization from band splitting generates magnetic moments that depend sensitively on electronic states at the Fermi level ($E_F$) [7]. Moreover, itinerancy also tends to drive magnetic frustration [8]. Thus, achieving high–$T_N$ metallic AFM materials is challenging due to the stringent balance required among atomic, spin, and electronic degrees of freedom; hence, most 2D antiferromagnets are either insulating or limited to low $T_N$.

Amid these constraints, the $Fe_NGeTe_2$ (FGT; $N$ = 3, 4, and 5) [9–11] family emerges as a rare room-temperature itinerant vdW ferromagnet, which also hosts high–$T_N$ itinerant antiferromagnetism via partial Co substitution [12,13]. The delocalized electrons likely contribute to their high ordering temperature [14]. However, this raises a fundamental question: *why is high ordering temperature not universal among all vdW itinerant magnets?* This alludes to an additional mechanism that must operate in FGT to stabilize its unusually high $T_C$ or $T_N$. Indeed, these materials display enhanced electronic correlation [15] and even heavy fermion behavior [16], reminiscent of *f*–electron systems [17], with these correlated states strongly coupled with their magnetic order [15]. As a result, although FGT is a prototypical itinerant magnet, its magnetism extends beyond the Stoner model, where partially screened local moments also contribute to the high $T_C$ [18,19].

In this letter, we study a new 35% Co-substituted $Fe_4GeTe_2$, $(Fe_{0.65}Co_{0.35})_4GeTe_2$ (FCGT), and identify the microscopic origin of its high $T_N$ (≈ 250 K) itinerant AFM state. Co substitution induces layer-selective Fe/Co ordering, resulting in a robust AFM structure. The AFM interactions are further stabilized by the strongly correlated Co 3*d* bands at $E_F$, giving rise to an itinerant-localized duality. Our results highlight the synergistic interplay of crystal, magnetic, and electronic structures in stabilizing a rare near–room–temperature antiferromagnetism in a vdW metal.

Table I [Supplemental Information (SI) [20]] summarizes representative vdW antiferromagnets, highlighting the scarcity of layered high–$T_N$ metals. Co–doped $Fe_4GeTe_2$ (F4GT) [13] and $Fe_5GeTe_2$ (F5GT) [12] exemplify such rare material candidates. Here, we focus on pristine and Co–doped F4GT because F5GT exhibits a complex sample–dependent magnetic behavior due to the presence of a mesoscale Fe–deficient secondary phase [21]. Moreover, the microscopic role of Co substitution in F4GT remains unclear, particularly how Co 3*d*–electrons mediate magnetic interactions and shape the electronic structure. In addition, a prior study also reported an unidentified magnetic phase at low temperature in Co–substituted F4GT [13]. The single crystals of these compounds (inset, Fig. S1) were synthesized using a chemical vapor transport method (details in SI [20]). The XRD spectra of these single crystals (Fig. S1) reproduce the characteristic (00*L*) reflections of pristine F4GT [9], indicating that FCGT retains the F4GT–type layered structure. The reduced (~1.1%) *c*–axis lattice parameter compared to that of F4GT ($c \approx 29.09°$) is consistent with smaller atomic radii of Co relative to Fe, confirming successful Co substitution. However, the similar atomic numbers of Fe and Co hinder a precise determination of their site occupancies [12]. To circumvent this problem, we performed density functional theory (DFT) calculations with different site–substituted configurations (Fig. S2 in SI [20]). These calculations reveal that the energetically most stable configuration places all four Co atoms within a single Fe layer (Fig. 1b), rather than distributing them across multiple layers. This analysis indicates a rare site–selective Co substitution in the Fe sub–lattice.

As shown in Fig. 2a, the temperature–dependent susceptibility [$\chi(T)$] with the in-plane ($H//ab$) and out-of-plane ($H//c$) magnetic fields of $\mu_0 H$ = 0.1 T displays a transition around $T_C \approx$ 270 K for F4GT, with clear irreversibility between the zero–field–cooling (ZFC) and field–cooling (FC) measurements. The field dependence of magnetization [$M$(H)] over the temperature range $T$ = 400–10 K shows saturation behavior below $T_C$ (Fig. 2c), with a lower saturation field ($H_S$) for $H//ab$ than $H//c$ near $T_C$ (Fig. 2c), indicating *ab*-plane as the magnetic easy axis at high temperature in F4GT. In contrast, FCGT exhibits AFM ground state, evidenced by a peak around $T \approx 250$ K and the absence of ZFC–FC splitting [Fig. 2b]. This $T_N \approx 250$ K is corroborated by a sharp anomaly in the specific–heat measurement (Fig. S3a). The out–of–plane $M$(H) curve reveals a metamagnetic spin-flop (SF) transition (blue dashed lines in Fig. 2c), whereas no such feature appears for the $H//ab$ field (Fig. S5b), indicating the AFM state with moments preferentially aligning along the *c*-axis. Within the FM configuration, the Co density of states (DOS) appears completely different than the Fe DOS at $E_F$. Interestingly, the overall DOS spectra of FCGT for the FM and AFM configurations appear similar (Fig. S5). However, integrating the DOS spectra up to $E_F$ captures subtle differences in the electronic occupation between the two magnetic states. Such a small rearrangement in the DOS leads to the slight lowering of the total energy by ~0.9 meV/magnetic atom in the AFM configuration, thereby confirming the stability of the experimentally observed AFM order.

Further, calculations show that flipping intralayer Co spins only slightly reduces the exchange energy (≈ –0.9 meV/magnetic atom), suggesting that they can rotate at marginal energy cost without disrupting the overall AFM order. Experimentally, cooling below $T_N$ results in a crossover between in-plane ($\chi_{ab}$) and out-of-plane ($\chi_c$) susceptibility near ~160 K (blue arrow; Fig. 2b). An anomaly in $\chi_{ab}$ is also observed around ~110 K in F4GT (blue arrow; Fig. 2a), accompanied by a reversal of the $H_S$ anisotropy (Fig. 2d), signaling a switching of the easy axis at temperature $T_{SR} \approx 110$ K (Fig. 1a). In FCGT, however, in–plane and out–of–plane $H_S$ remain nearly identical across all temperatures (Fig. 2d). The out–of–plane *M*(H) curves show hysteresis near the SF transition below ~160 K (Fig. 2c), corroborated by the separation of peaks in d*M*/d*H* between increasing and decreasing field sweeps (Fig. S6a). These features collectively point to the emergence of a weak FM component arising from spin canting. A similar behavior is seen in the compound $Fe_{1-x}Co_xCl_2$ at comparable Fe and Co concentrations ($0.2 < x < 0.5$), where competing single–ion anisotropies are known to induce such canting [22,23]. Consistently, the growing discrepancy between $H_{SF}$ and $H_S$ below $T_{SR}$ (Fig. S2c) reflects the larger fields required to align canted moments along the *c*–axis.

DFT calculations demonstrate that the Co atoms adopt an intralayer AFM order, generating canted Co magnetic moments (1.2 $\mu_B$ and 0.3 $\mu_B$). The same calculations incorporating Hubbard parameters ($U_{eff}$ = 4 eV) change these values to 2.2 $\mu_B$ and 1.5 $\mu_B$. The Fe moments from both sets of calculations, on the other hand, align ferromagnetically within layers but antiferromagnetically across them, producing competing in–plane and out–of–plane easy magnetization directions. However, the total energy for *ab*-plane magnetization is lower by 0.89 meV per magnetic atom than *c*-axis magnetization, confirming the experimentally observed *ab*-plane ground state in FCGT. This result is in a complete contrast to that of the *c*-axis FM ground state in F4GT (Fig. 1a). Such a unique layer-selective AFM structure in FCGT also mitigates magnetic frustration by separating ions with distinct in-plane exchange tendencies (FM for Fe and AFM for Co). Consequently, competing phases such as the spin-glass states reported in $TaFe_{1.14}Te_3$ [24] and $Fe_{1/3}NbS_2$ [25] are avoided in FCGT, as manifested by the absence of ZFC–FC irreversibility in $\chi_{ab}$ and $\chi_c$ (Fig. 2b) and glassy hysteresis in $M$(H) (Fig. 2c and S5b). This likely plays a key role behind substantially higher $T_N$ of FCGT compared to other vdW AFM metals (Table I).

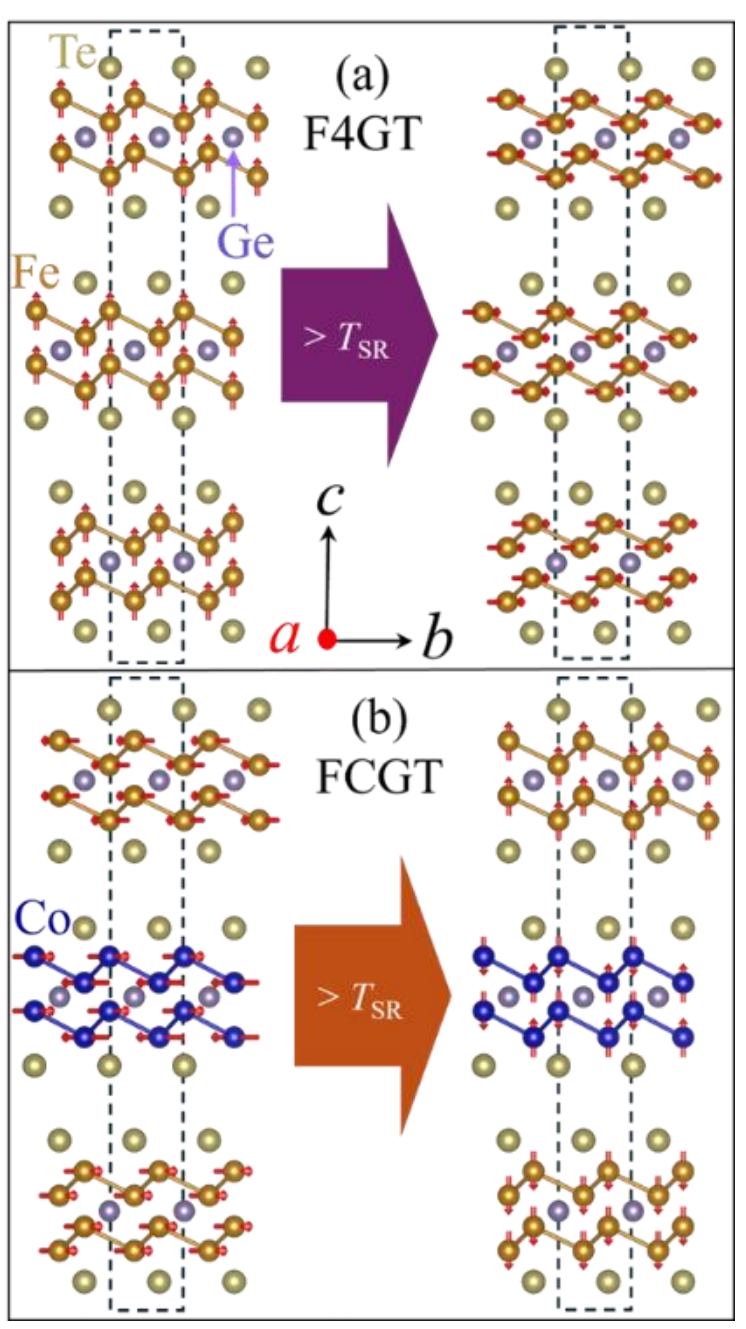

***FIG. 1.*** *Crystal and magnetic structures of (a) F4GT and (b) FCGT, both crystallizing in the rhombohedral structure (*$R\bar{3}m$ *space group) with the a, b, and c axes indicated. Left and right panels depict the low- and high-temperature magnetic states below and above* $T_{SR}$*, respectively.*

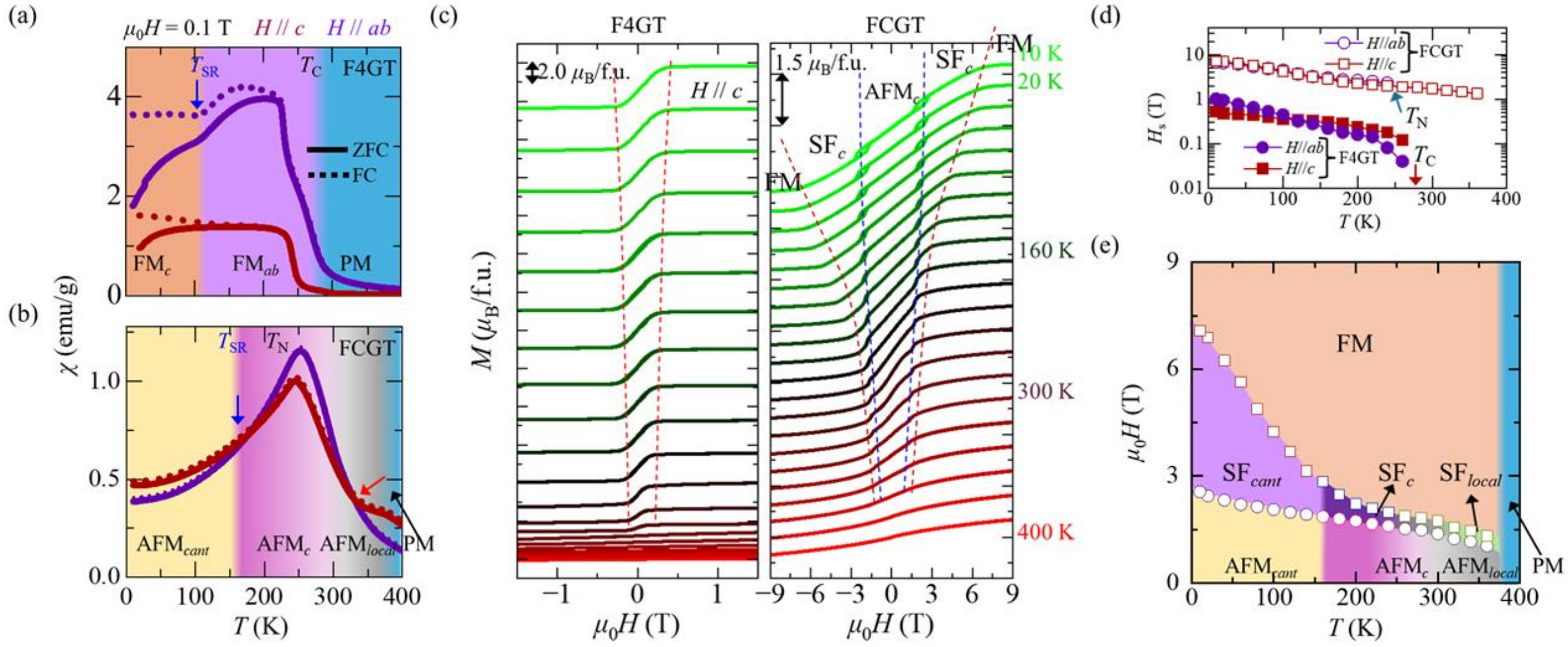

***FIG. 2.*** *Temperature dependence of susceptibility [χ(T)] measured under H//ab and H//c magnetic fields of $\mu_0 H$ = 0.1 T for (a) F4GT and (b) FCGT. Solid and dotted lines correspond to ZFC and FC measurements, respectively. The blue arrows in (a) and (b) indicate $T_{SR}$. The different colored regions in Figs.a and b denote distinct magnetic phases. Field dependence of magnetization [M(H)] for F4GT under H//ab field from T = 400 K to 10 K (c) F4GT and (d) FCGT. The red dashed line represents moment saturation, whereas the blue lines in Fig. 2d denote AFM-to-SF transition. (e) Temperature dependence of the in-plane and out-of-plane saturation fields ($H_S$) for both samples. (e) Magnetic phase diagram of FCGT as a function of temperature and magnetic field, which summarizes the evolution of magnetic states, including PM, $AFM_{local}$, $AFM_c$, and $AFM_{cant}$ phases. These states (except the PM state) exhibit field-induced transition to corresponding spin-flop phases ($SF_{local}$, $SF_c$, and $SF_{cant}$ phases) followed by the FM phase at high fields. All the symbols denote experimentally determined phase boundaries from χ(T) and M(H) data under H//c.*

The evolution of magnetic ordering below $T \leq 400$ K and field up to $\mu_0 H \leq 9$ T for FCGT is summarized in a phase diagram (Fig. 2e), established by precisely identifying the critical fields and temperatures for all temperature- and field-driven magnetic transitions. Such a rich sequence of magnetic states within this *T*-*H* phase space is unique to FCGT. For example, magnetic saturation in vdW antiferromagnets like $TaFe_{1.14}Te_3$ [24] and $AgCrSe_2$ [26] occurs under ultra–high fields above 20 T. The tunable magnetism in FCGT originates from the intrinsic flexibility of Co moments, consistent with the above–mentioned small FM–AFM energy difference, comparable to the Zeeman energy at the saturation field $H_S \approx 7$ T (~0.4 meV) at 10 K.

The Co substitution also has a pronounced impact on electrical transport. As seen in Fig. 3a, the longitudinal resistivity [$\rho(T)$] for F4GT exhibits overall metallic temperature–dependence, displaying clear slope changes near $T_C$ and $T_{SR}$, corresponding to the critical temperatures identified from the susceptibility data in Fig. 2a. In contrast, FCGT exhibits a weak increase (by a factor of ~1.3) on lowering the temperature from 300 to 10 K, with a broad hump near $T_N$ (Fig. 3b). This increase in $\rho(T)$ is much smaller than for vdW AFM semiconductors [27,28] but comparable to vdW AFM metals [24,29], which suggests metallic–like transport despite marginally negative d$\rho$/d$T$. To clarify this anomaly, we analyze DFT [30–33] results for FCGT, which reveal the metallic band structure with finite DOS at the $E_F$ (Fig. S4). Electronic states near the $E_F$ are mostly dominated by 3$d$–states of Co, with much smaller contributions from 3$d$– Fe, Ge 4$p$–, and Te 5$p$–states. The atom–projected DOS exhibits a valley–like minimum at $E_F$ for Fe, Ge, and Te bands (Fig. S4b), reflecting $p$–$d$ hybridization, which primarily drives the itinerant behavior of the system. In contrast, the Co–projected DOS shows opposite tendency with comparatively narrow DOS peaks in the occupied states and a quasi–peak at the $E_F$, indicating partially localized Co–3$d$ states. The markedly opposite behavior of Fe and Co 3$d$ bands near the $E_F$ in FCGT correlates with Co–substituted $Fe_3GeTe_2$ at comparable doping levels [34]. Reported similar DOS features near $E_F$, despite not fully isolating Fe and Co contributions, may suggest a generic behavior in Co-substituted FGT systems.

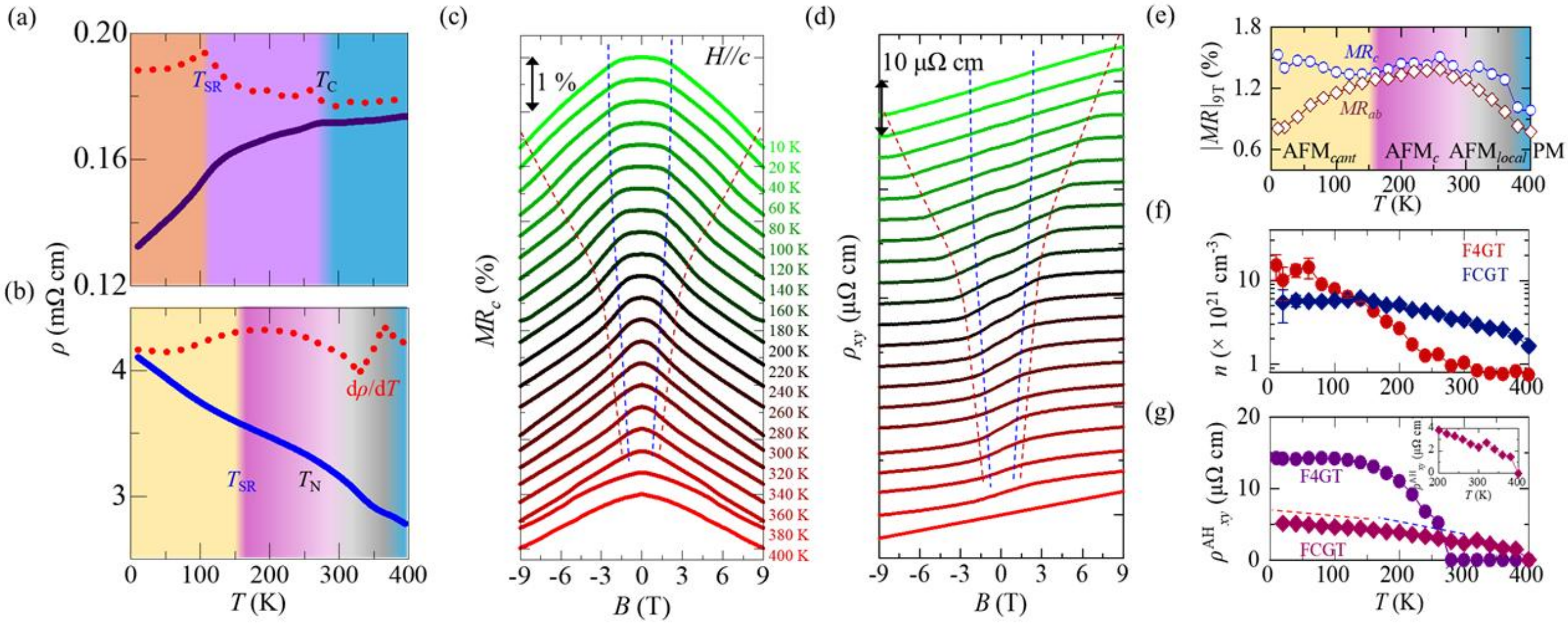

***FIG. 3.*** *Temperature dependence of the longitudinal resistivity for (a) F4GT and (b) FCGT, with red curves showing the corresponding derivatives dρ/dT. (c) Field dependence of Normalized magnetoresistance $MR_c$ of FCGT at different temperatures from T = 400 K to 10 K under H//c field.*

*(d) Field dependence of Hall resistivity of FCGT within the same temperature range. The blue and red dashed lines in Figs. c and d represent AFM–to–SF and SF–to–FM transitions, respectively. (e) Temperature dependence of the magnitude of $MR_c$ and $MR_{ab}$ at 9 T field for FCGT. The three different colored regions in Fig. 3e represent different magnetic states. Temperature dependence of (f) carrier density (n), and (g) anomalous Hall resistivity $\rho^{AH}_{xy}$ of both F4GT and FCGT samples. The inset highlights the evolution of $\rho^{AH}_{xy}$ in the high temperature regime of FCGT. All dashed lines in Fig. 3g are guides to the eye.*

We also extracted the DOS at the $E_F$ ($D(E_F)$) from the specific–heat (C(T)) measurements (detailed discussed in SI [20]). A linear fit to $C/T = \gamma + \beta T^2$ yields $\gamma$ = 86.5 mJ/ (mol·$K^2$), corresponding to a $D(E_F) \approx 36$ states/(eV·f.u.) via the relation $\gamma = \frac{\pi^2 K_B^2 D(E_F)}{3}$. The *ab initio* calculated $D(E_F)$ value (36.84 states / eV . f.u.) closely agrees with the experiment. As shown in SI [20], the Fermi level lies around the Co DOS peak. The unusually large $\gamma$ value, therefore, is ascribed to strong electronic correlations [15,16], which likely stem from renormalized Co–3*d* bands near $E_F$. This points to a dual electronic character in FCGT, reminiscent of the itinerant–localized duality of 3*d* electrons [18]. In this picture, quasi–localized Co–3*d* states introduce strong spin–dependent scattering, likely contributing to the non–metallic transport behavior (Fig. 3b. Similarly, such a large $D(E_F)$ arising from renormalized bands can also strengthen magnetic order in a mixed itinerant–localized system [35].

A similar coexistence of itinerant and localized electronic states has been suggested in an FGT system [18]. Here, we show that their interplay stabilizes robust local moments and sustains strong magnetic correlations even after the loss of long–range magnetic order, providing a key mechanism for the high $T_N$ in FCGT. Above $T_N$, an additional crossover between $\chi_{ab}$ and $\chi_c$ occurs near ~320 K (red arrow, Fig. 2b), accompanied by a dip in d$\rho$/d$T$ (Fig. 3b) at the same temperature. On further heating, a broad anomaly in $\chi_c$ (Fig. 2b) and a subtle kink in $\rho$(T), seen as a peak in d$\rho$/d$T$ (Fig. 3b), emerge around ~360 K. Consistently, the SF transition in out–of–plane $M$(H) (Fig. 2c), manifested as a peak in d$M$/d$H$ (Fig. S7a), also survives up to this temperature. This behavior aligns with the *c*–axis anisotropy imposed by Co moments, as discussed above. In contrast, pristine F4GT shows linear $M$(H) with no saturation behavior immediately above $T_C$ (Figs. 2c and S6a), underscoring the role of localized Co moments in pushing magnetic correlations well beyond $T_N$. We therefore assign the temperature regime between $AFM_c$ and 360 K as $AFM_{local}$. This regime follows $\chi^{-1} \propto (T-T_C)^{1-\lambda}$ behavior with $\lambda = 0.66$ (details in SI; Fig. S8), pointing to a Griffiths-like magnetic phase [36].

Magnetotransport response further provides a complementary probe of magnetic order via spin-dependent scattering. Since high-temperature magnetic correlations are more evident in $\chi_c$ than $\chi_{ab}$, we measured $MR_c$ (Fig. 3c) and $MR_{ab}$ (Fig. S9) under $H \perp I$ and $H \parallel I$ fields, respectively, to directly compare the responses of the out–of–plane and in–plane moment. FCGT exhibits negative $MR_c$ and $MR_{ab}$ across $T$ = 10–400 K, consistent with field-induced suppression of spin scattering. The $MR_c$ mirrors the out–of–plane $M$(H) traces, revealing distinct signatures at the AFM–to–SF (blue dashed lines) and SF–to–FM (red dashed lines) transitions. In the $AFM_c$ phase, both responses are comparable and weakly temperature–dependent (Fig. 3e). By contrast, below $T_{SR}$, $MR_c$ remains nearly constant at around ~1.3–1.5 % down to 10 K, whereas $MR_{ab}$ reduces to ~0.8 % at 10 K. This emerging anisotropy likely arises from moment canting at low temperature. A similar MR anisotropy is also observed in the $AFM_{local}$ regime with $MR_c > MR_{ab}$, where $MR_c$ shows a slight enhancement near the $AFM_c$–$AFM_{local}$ phase boundary before dropping upon entering the PM state. On the other hand, $MR_{ab}$ decreases monotonically above $T_N$.

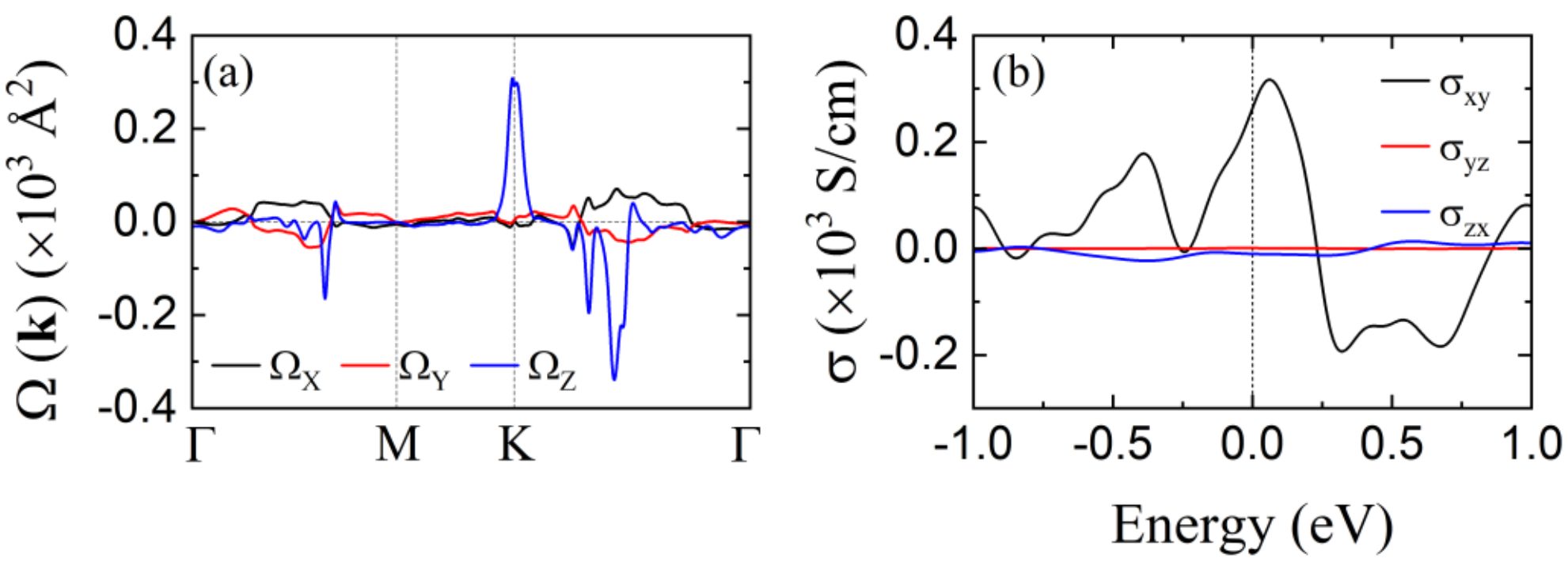

***FIG. 4.*** *(a) Berry curvature $\Omega(\mathbf{k})$ (in $10^3$ Å$^2$) along the high-symmetry path and (b) anomalous Hall conductivity components $\sigma_{xy}$, $\sigma_{yz}$, and $\sigma_{zx}$ (in $10^3$ S/cm) near the $E_F$ as a function of energy.*

The Hall resistivity ($\rho_{xy}$) of FCGT crystals exhibits a non–linear field dependence (Fig. 3d), following distinct features seen in the $MR_c$ near the $H_{SF}$ and $H_S$ fields. Further, a linear relationship between $\sigma^{AH}_{xy}$ and $M_S$ is obtained above $T_{SR}$ (Fig. S12a), suggesting an intrinsic magnetism-driven anomalous Hall effect (AHE) [37]. Such linear scaling is also a hallmark of a Berry-curvature-driven AHE [37], which is suggested by $\sigma^A_{xy}$–$\sigma_{xx}$ scaling behavior (Fig. S12b; details in SI [20]).

This is also supported by *ab initio* calculations (Fig. 4). A pronounced peak in the Berry curvature $\Omega^z(\mathbf{k})$ suggests a topologically non-trivial band structure and supports robust edge states [30]. The Berry curvature acts as a fictitious magnetic field in momentum space, directly contributing to the AHC and exhibiting a strong energy dependence in the vicinity of the $E_F$ (Fig. 4). Experimental results show that the $\rho^{AH}_{xy}$ decreases gradually from 10 K to $T_{SR}$, followed by a more rapid suppression up to the $AFM_c$–$AFM_{local}$ phase boundary (denoted by the dashed lines), reflecting the strong sensitivity of AHE to changes in the underlying magnetic configuration. With further temperature increase, $\rho^{AH}_{xy}$ abruptly enhances before it again decreases up to ~360 K and completely vanishes above this temperature (inset, Fig. 3g). This is in stark contrast to the disappearance of AHE immediately above $T_C$ in F4GT (Figs. 3g and S11). Such high-temperature anomalies in $MR_c$ and $\rho^{AH}_{xy}$ correlate with the $\chi_c$ features in FCGT (Fig. 2b), confirming substantial magnetic fluctuations above $T_N$ originating from local Co moments.

Moreover, in FCGT, a dominant hole–type conduction with the carrier density $n \approx (1.64\pm0.08)$ to $\approx (6.18\pm0.11)\times10^{21}$ cm$^{-3}$ at a temperature $T$ = 10–400 K is obtained (Fig. 3f). On the other hand, the $n$ for F4GT span between $\approx (0.74\pm0.02)$ and $\approx (15.22 \pm 5.06) \times 10^{21}$ cm$^{-3}$ (Fig. 3f). For FCGT, $n$ shows no significant variation with $n_{10K}/n_{400K} \approx 3.8$, indicating the absence of a pronounced temperature–driven electronic band reconstruction. In contrast, F4GT exhibits a much larger variation, $n_{10K}/n_{400K} \approx 20.6$ (Fig. 3f), accompanied by a sharp decrease in $n$ above $T_C$, consistent with Stoner–like exchange–induced band splitting. In fact, a substantial modification of the electronic structure across $T_N$ is reported in vdW antiferromagnets, FeTe [38] and $TaFe_{1.14}Te_3$ [24]. These results indicate a unique interplay between magnetism and electronic structure in FCGT, highlighting a crucial role of Co moments and quasi–localized Co 3*d*–states in stabilizing its high–$T_N$ AFM order.

In conclusion, our study demonstrates that the high $T_N$ in $(Fe_{0.65}Co_{0.35})_4GeTe_2$ compound arises from the synergy between layer–selective Fe and Co substitution, robust AFM interactions primarily driven by Co moments, and the itinerant–localized duality of 3*d*–electrons. This combination generates strong magnetic correlations well above $T_N$, while also stabilizing spin canting at low temperature with a possibility of a non–trivial Berry curvature. These insights highlight a pathway for the discovery of a new room–temperature vdW itinerant antiferromagnet and a promising platform for AFM spintronics.

## Acknowledgements

This work at Morgan State University has been funded by the Department of Defense through grant # W911NF2120213. The theoretical contribution of D.P. & H. P. was supported as part of the Center for Energy Efficient Magnonics, an Energy Frontier Research Center funded by the U.S. Department of Energy, Office of Science, Basic Energy Sciences under contract #DEAC02-76SF00515. The specific-heat capacity measurement at the University of Maryland was supported by the National Science Foundation under the Division of Materials Research Grant NSF-DMR 2514293. Identification of commercial equipment is for information purposes only and does not imply endorsement by NIST.